\documentstyle[12pt,epsfig]{article}
\newcommand{\PR}[1]{{\it Phys.\ Rev.}\ {\bf #1}}

\newcommand{\simorder}{\raisebox{-4pt}{$\, \stackrel{\textstyle >}{\sim} \,$}} 

\newcommand{\la}{\lambda}
\newcommand{\La}{\Lambda}
\newcommand{\bfk}{\mbox{\boldmath $k$}}
\newcommand{\pup}{p^\uparrow}
\newcommand{\pdown}{p^\downarrow} 
\newcommand{\qup}{q^\uparrow} 
\newcommand{\qdown}{q^\downarrow} 
\newcommand{\aup}{a^\uparrow} 
\newcommand{\adown}{a^\downarrow} 
\newcommand{\bup}{b^\uparrow} 
\newcommand{\bdown}{b^\downarrow} 
\newcommand{\cp}{c^\uparrow} 
\newcommand{\cdown}{c^\downarrow} 

\newcommand{\bfp}{\mbox{\boldmath $p$}}
\newcommand{\bfP}{\mbox{\boldmath $P$}} 
\newcommand{\bfs}{\mbox{\boldmath $s$}} 
\newcommand{\bfz}{\mbox{\boldmath $z$}} 
\newcommand{\bfy}{\mbox{\boldmath $y$}} 
\newcommand{\Lup}{\Lambda^\uparrow} 
\newcommand{\Ldown}{\Lambda^\downarrow} 
\newcommand{\hup}{h^\uparrow} 
\newcommand{\hdown}{h^\downarrow} 
 
\newcommand{\Nup}{N^\uparrow} 
\newcommand{\nd}{\noindent}
\newcommand{\be}{\begin{equation}}
\newcommand{\ee}{\end{equation}}
\newcommand{\bea}{\begin{eqnarray}}
\newcommand{\eea}{\end{eqnarray}}
\begin{document}
\begin{center}
{\bf Single spin asymmetries in QCD\footnote{Lectures delivered 
at the Advanced Study Institute, "Symmetries and Spin" - Praha-SPIN-2001,
Prague, July 15 - July 28, 2001}}   
\vskip 0.4cm
{\sf Mauro Anselmino}
\vskip 0.6cm
{\it Dipartimento di Fisica Teorica, Universit\`a di Torino and \\
INFN, Sezione di Torino, Via P. Giuria 1, I-10125 Torino, Italy}
\end{center}

\vspace{0.6cm}

\begin{abstract}
Measurements of single transverse spin asymmetries in high energy 
inclusive processes have always shown unexpected and challenging results.
Several cases are considered and discussed within a QCD approach 
which couples perturbative dynamics to new non perturbative partonic
information; the aim is that of developing a consistent phenomenological 
description of these unusual single spin phenomena, based on a 
generalized QCD factorization scheme.   
\end{abstract}

\vspace{0.6cm}
\nd
{\bf 1. Introduction and notations}     
\vskip 8pt

Spin measurements always test a theory at a much deeper level than 
unpolarized quantities, spin being an intrinsically relativistic and 
quantum mechanical aspect of particle interactions, which can rarely 
be described by classical models, contrary to what happens, for example, 
for unpolarized cross-sections. 

Among spin measurements, single spin asymmetries in high energy processes 
involving nucleons are even more peculiar and interesting: many of them are 
observed to be large and sizeable, yet they do not correspond to a similar
asymmetry in the elementary interactions among the nucleon constituents,
and their origins is related to new and subtle non perturbative features
of parton hadronization and/or distribution. It is a typical case in which
the simplicity of the short distance physics is not transmitted to the
long distance observables.

I consider here several of these cases, arguing in favour of a possible 
consistent interpretation and phenomenological description; the framework 
of these lectures is pQCD and the factorization scheme, according to which
the differential cross-sections for processes involving high energy scales    
can be written as a convolution of hard, perturbative and short distance
dynamics with soft, non perturbative and long distance contributions. 
The former is known from QED, QCD or the Standard Model of electroweak
interactions while the latter have to be taken from experiments,
in form of parton distribution and/or fragmentation functions: however, QCD 
has succeeded in giving us the energy evolution of these universal partonic 
properties, so that only measurements at an initial scale in one process
are indeed necessary and true predictions at any other scale and for different
processes are possible and successful. 

To set the notations, the differential cross-sections for high energy and 
large momentum transfer inclusive processes like $\ell \, N \to \ell \, X$ 
(DIS) or $N \, N \to h \, X$, where $\ell$ is a lepton, $N$ a nucleon (proton 
or neutron), $h$ an observed final hadron and $X$ all the non detected final 
particles, are given in the QCD factorization scheme by \cite{[1]}:
\be
d\sigma^{\ell N \to \ell X} = \sum_a \int_0^1 \!\! dx \> f_{a/N}(x, Q^2) \>
d\hat\sigma^{\ell a \to \ell \dots } \label{unpdis}
\ee
and
\be
d\sigma^{N N \to h X} = \sum_{a,b,c} \int \!\! dx_1 \, dx_2 \> 
f_{a/N}(x_1, Q^2) \> f_{b/N}(x_2, Q^2) \>
d\hat\sigma^{a b \to c \dots } \> D_{h/c}(z,Q^2) \label{unpnn}
\ee
where $f_{a/N}(x,Q^2)$ is the unpolarized distribution function of parton 
$a$, which carries a longitudinal momentum fraction $x$, inside a nucleon 
$N$, probed at small distances, related to a large squared four-momentum 
transfer $Q^2$. Similarly, $D_{h/c}(z,Q^2)$ is the fragmentation function 
of parton $c$ into a hadron $h$, which carries a longitudinal momentum 
fraction $z$, seen at a scale $Q^2$. The $d\hat\sigma$'s are the elementary 
hard cross-sections given by QED, QCD or the Standard Model, according to 
the type of interactions contributing.

Similar formulae hold for the polarized case: for example, with leptons and 
nucleons longitudinally polarized one has new information from double spin  
asymetries in DIS    
\be
d\sigma^{++} - d\sigma^{+-}
= \sum_a \int_0^1 \!\! dx \> \Delta f_{a/N}(x, Q^2) \> \left[
d\hat\sigma^{++} - d\hat\sigma^{+-} \right] \label{poldis}  
\ee
where the apices refer to the lepton and nucleon (or lepton and parton, 
in $d\hat\sigma$) helicities and $\Delta f_{a/N}(x,Q^2)$ is the helicity 
distribution of parton $a$, also indicated by $\Delta q$ for quarks of 
flavour $q$ and $\Delta g$ for gluons:
\be
\Delta f_{a/N}(x,Q^2) \equiv f_{a_+/N_+}(x,Q^2) - f_{a_-/N_+}(x,Q^2) \>.
\label{delq}
\ee

All distribution functions obey the well known QCD evolution in $Q^2$ 
\cite{[1]}, so that measurements of the cross-sections at a certain scale 
$Q_0^2$ in DIS can be used to obtain information on the parton distributions 
at the same scale, which, once evolved, allow predictions and test of the 
theory at other scales and in other processes.  

\vskip 12pt 
{\bf 1.1 Elementary spin dynamics}
\vskip 8pt

The partonic elementary interactions entering Eqs. 
(\ref{unpdis})-(\ref{poldis}) have a very simple spin dynamics.
The spinors for massless quarks and antiquarks with helicity
$\la = \pm 1$, have simple projector operators:
\bea
\gamma_5 \, u_\la = \la \, u_\la \quad\quad\quad &&
\frac{1 + \la \,\gamma_5}{2} \, u_\la = u_\la \label{pr1} \\
\bar u_\la \, \gamma_5 = -\la \, \bar u_\la \quad\quad\quad &&
\bar u_\la \, \frac{1 - \la \,\gamma_5}{2} = \bar u_\la \>. 
\label{pr2} 
\eea
Let us consider a quark with helicity $\la$ interacting with any number 
of photons, gluons, $Z$'s or $W$'s and resulting in a final quark with 
helicity $\la'$: it contributes to the scattering amplitude with an 
expression of the kind
\be
M \sim \bar u_{\la'} \, \Gamma \, u_\la
\ee
where $\Gamma$ can always be written as the product of an {\it odd} number
of $\gamma$ matrices, resulting from vertices and propagators.
Eqs. (\ref{pr1}) and (\ref{pr2}) then imply:
\bea
M &\sim& \bar u_{\la'} \left( \frac{1 - \la'\,\gamma_5}{2} \right)
\Gamma \left( \frac{1 + \la \,\gamma_5}{2} \right) u_\la 
\nonumber \\
&\sim& \bar u_{\la'} (1 - \la'\,\gamma_5)\,(1 - \la \,\gamma_5)
\, \Gamma \, u_\la \sim \delta_{\la\la'} \nonumber
\eea
which shows the well known {\it helicity conservation} of strong and
electromagnetic interactions. 

Analogously, one can show that a quark and an antiquark {\it can annihilate 
only if they have opposite helicities.}

The above results can only be corrected by terms proportional to the ratio
of the quark mass to its energy, which are negligible at high energies.   

This dynamical simplicity makes spin effects in elementary interactions
easily computable; in particular it makes impossible to have transverse 
single spin asymmetries in, say, quark-quark or quark-gluon interactions.

Let us consider, for example, the elastic scattering of two quaks of 
different flavour, $q \, q' \to q \, q'$. All observables can be expressed 
in terms of 6 independent helicity amplitudes, 
$M_{\la^{\prime}_{q},\la^{\prime}_{q'};\la_{q}\la_{q'}}$,
which are usually denoted as:
\bea
&& M_{++;++} \equiv \Phi_1 \quad\quad    
   M_{--;++} \equiv \Phi_2 \quad\quad 
   M_{+-;+-} \equiv \Phi_3 \quad\quad \nonumber \\
&& M_{-+;+-} \equiv \Phi_4 \quad\quad    
   M_{-+;++} \equiv \Phi_5 \quad\quad 
   M_{++;+-} \equiv \Phi_6 \>. \nonumber
\eea
For parity conserving interactions the only single spin asymmetries allowed 
are for polarizations orthogonal to the scattering plane: if we have
the two initial quarks unpolarized and look at this transverse polarization
of one of the final quarks, say $q$, we can have different cross-sections
$\sigma(\uparrow) \not= \sigma(\downarrow)$. One can show that \cite{bls}
\be
P_q \equiv \frac{\sigma(\uparrow) - \sigma(\downarrow)}
                {\sigma(\uparrow) + \sigma(\downarrow)}
\sim {\rm Im} \left[ \Phi_6\,(\Phi_1 + \Phi_3)^* - 
                \Phi_5\,(\Phi_2 - \Phi_4)^* \right] \label{polq} 
\ee
which can be non zero only if some of the single helicity flip amplitudes
is non zero and different amplitudes have relative phases. The first 
condition implies a factor $m_q/E_q$ (quark mass over its energy) and the 
second requires considering higher order contributions (as lowest order 
there are no relative phases). Altogether one has 
\be
P_q \simeq \frac{m_q}{E_q}\,\alpha_s \>, \label{parpol}
\ee
where $\alpha_s$ is the strong coupling constant. This makes single spin
asymmetries in the partonic interactions entirely negligible. 

\vskip 12pt    
{\bf 1.2 Hadronic single spin asymmetries}
\vskip 8pt

Helicity is a good and natural quantum number at the partonic level,
rising hopes for simple spin observables also in the hadronic world;
is this true? 
  
The answer is definitely no, at least up to the energy ranges where
data are available. Whether or not the partonic simplicity will appear in 
hadronic spin observables at higher energies remains to be seen; for
the moment, let us look at the data, and let us try to understand them.   

I shall consider the following single transverse spin phenomena:
\begin{itemize}

\item
Polarization of hyperons (mainly $\La$'s) produced in unpolarized
hadronic processes, $p\,N \to \Lup \, X$ \cite{lamdata}; this is a well 
known puzzle, dating from almost 20 years ago: how unpolarized initial 
hadrons, presumably containing unpolarized partons, might give origin, 
when colliding, to polarized hadrons \cite{lamth}. The observed polarization 
can easily reach magnitudes of 20\%, while the analogous observable in 
elementary QCD dynamics, Eq. (\ref{parpol}), is negligible. 

Hyperons are considered, rather than other baryons, for the simple reason
that it is possible to measure their polarization by studying the angular 
distribution of their weak decays. The $\La$, for example, decays
weakly into a pion and a proton, $\La \to p\,\pi^-$, and the angular 
distribution of the proton in the $\La$ rest frame is given by
\be
W(\theta,\varphi) = \frac{1}{4\pi}\,(1 + \alpha \, \bfP_\La \cdot
\hat{\bfp}) \label{plam}
\ee
where $\alpha \simeq 0.64$ and $\hat{\bfp}$ is the unit vector along the 
proton direction. 
  
\item
Another amazing piece of experimental information comes from E704 
experiments at FERMILAB \cite{e704}, with the collision of a 
transversely polarized proton beam off an unpolarized target, and the
observation of inclusively produced pions, $\pup \, p \to \pi \, X$.
The number of produced pions, in certain kinematical ranges, depends 
strongly on the ``up'' or ``down'' proton polarization, again showing a 
hadronic single spin asymmetry absent in the quark and gluon interactions. 

\item
Finally, I consider the azimuthal asymmetry of pions produced in 
semi-inclu\-si\-ve deep inelastic scattering (SIDIS), with unpolarized leptons 
and polarized protons, $\ell \, \pup \to \ell \, \pi\, X$. When looking at the 
process in the center of mass frame of $\gamma^*$-$\pup$, with the proton
transversely polarized, the number of produced pion depends on the azimuthal 
angle of the pion around the $\gamma^*$ direction \cite{herm,smc}.  

\end{itemize}

In the sequel I discuss at length the above phenomena, 
trying to give a consistent and unified description in a 
generalized version of the QCD factorization scheme, which allows for
intrinsic transverse motion of partons inside hadrons, and of hadrons
relatively to fragmenting partons; this, as we shall see, adds new 
possibilities of spin effects, absent for collinear configurations.

\vskip 12pt \nd
{\bf 2. Origins of hadronic single spin asymmetries}
\vskip 8pt

We start by discussing the transverse single spin asymmetries observed in 
$\pup \, p \to \pi \, X$, which allow to introduce the modified formalism
to be used in general for the single spin effects.
The process is described in the $p$-$p$ center of mass frame: the protons
move along the $z$-axis, with the polarized beam along $+\hat{\bfz}$; 
$xz$ is the scattering plane; the pion has momentum $\bfp_\pi = \bfp_L + 
\bfp_T$; $p_L = x_F \sqrt s /2$ and $p_T$ are respectively the magnitude 
of the longitudinal and transverse components; the proton spin is either 
$\uparrow$ (along $+\hat{\bfy}$) or $\downarrow$ (along $-\hat{\bfy}$).

The observed asymmetry is
\be
A_N(x_F, p_T) \equiv 
\frac{d\sigma^\uparrow(x_F, \bfp_T) - d\sigma^\downarrow(x_F, \bfp_T)}
     {d\sigma^\uparrow(x_F, \bfp_T) + d\sigma^\downarrow(x_F, \bfp_T)} 
\label{spa}
\ee
where $d\sigma$ is the invariant cross-section $E_\pi\,d^3\sigma/d^3\bfp_\pi$.
Rotational invariance implies $d\sigma^\downarrow(x_F, \bfp_T) = 
d\sigma^\uparrow(x_F, -\bfp_T)$ so that $A_N$ is ofted referred to as 
{\it left-right asymmetry.} 

According to the QCD factorization theorem at leading twist, with collinear 
configurations, the cross-section for the high energy and large $p_T$
unpolarized process $p \, p \to \pi \, X$ is given by Eq. (\ref{unpnn}), which 
can be written shortly as: 
\be
d\sigma = \sum_{a,b,c} f_{a/p} \otimes f_{b/p} \otimes
d\hat\sigma^{ab \to c \dots} \otimes D_{\pi/c} \>.
\label{lt}
\ee

In such a scheme there is no possibility for single spin effects. There is 
no term in Eq. (\ref{lt}) which can have a single spin dependence: the 
collinear distribution and fragmentation processes (which can be visualized 
as a proton splitting into a parton + remnants, or a parton fragmenting
into a pion + remnants, {\it all being collinear}) cannot depend on a single 
spin due to rotational invariance and single spin asymmetries are negligible 
in elementary interactions. Eq. (\ref{lt}) cannot explain the large observed
effects \cite{e704} and this prompted a copious theoretical activity.

Among the attempted explanations of $A_N$ observed in E704 experiment
there are generalizations of the QCD factorization theorem with the
inclusion of higher twist correlation functions \cite{ekt,qs,kk}, or with the 
inclusion of intrinsic $\bfk_\perp$ and spin dependences in distribution
\cite{siv,noi1,bm,dan} and fragmentation \cite{col,bm,noi2,bl,suz} functions; 
there are also some semi-classical approaches based on the introduction of 
quark orbital angular momentum \cite{bztm,bzt}. I consider here only the 
approaches which are based on QCD dynamics, through a generalization of the 
factorization scheme. A review paper on the subject can be found in 
Ref. \cite{bzt}. I would like to stress that the introduction of parton 
transverse motion is known to be necessary also to describe the unpolarized 
cross-sections \cite{wang,apa}.  

\goodbreak
\vskip 12pt
{\bf 2.1 Higher twist parton correlations}
\nobreak
\vskip 8pt

In the approach of Ref. \cite{qs} Eq. (\ref{lt}) is generalized -- and proven
to hold -- with the introduction of higher twist contributions to distribution 
or fragmentation functions. Schematically it reads: 
\bea  
&& d\sigma^\uparrow - d\sigma^\downarrow = \sum_{a,b,c} 
\Bigl\{ \Phi_{a/p}^{(3)} \otimes f_{b/p} \otimes \hat H \otimes D_{\pi/c} 
\nonumber \\
&+& h_1^{a/p} \otimes \Phi_{b/p}^{(3)} \otimes \hat H' 
\otimes D_{\pi/c} + h_1^{a/p} \otimes f_{b/p} \otimes \hat H'' \otimes 
D_{\pi/c}^{(3)} \Bigr\}\>,
\label{htqs}
\eea 
where the $\Phi^{(3)}$'s, $D^{(3)}$'s, are higher twist partonic correlations 
(rather than parton distributions or fragmentations) and the $\hat H$'s denote
the elementary interactions. $h_1$ is the transversity distribution, defined
analogously to the helicity distribution of Eq. (\ref{delq}):  
\be
h_1^{a/N}(x,Q^2) \equiv f_{\aup/\Nup}(x,Q^2) - f_{\adown/\Nup}(x,Q^2) \>.
\label{h1}
\ee

The higher twist contributions are unknown, but some simple models can 
be introduced, for example \cite{qs} 
\bea
&& \!\!\!\!
\Phi^{(3)}_{a/p} \sim \int\frac{dy^-}{4\pi}\>  e^{ixp^+y^-} \langle p, \bfs_T|
\, \overline\psi_a(0) \gamma^+ \times \nonumber \\
&& \!\!\!\!\!\!\!\!\!\! \left[ \int \!\! dy_2^- \, 
\epsilon_{\rho\sigma\alpha\beta} 
\> s_T^\rho \, p^\alpha \, p'^{\beta} \, F^{\sigma+}(y_2)\right]
\psi_a(y^-) \, |p, \bfs_T \rangle 
=  k_a \> C \> f_{a/p} \>. \label{phi3}
\eea

The above contribution depends on the initial nucleon momenta $p$ and $p'$,
on the transverse proton spin $\bfs_T$ and on some external gluonic field
$F^{\mu\nu}$; it involves transverse degrees of freedom of the partons
and it differs from the usual definition of the distribution functions
$f_{a/p}$ only by the insertion of the term in squared brackets.
This is the reason for the last term in Eq. (\ref{phi3}), where $C$ is a 
dimensional parameter and $k_a$ is respectively $+1$ and $-1$ for $u$ and 
$d$ quarks.

Such a simple model can reproduce the main features of the data \cite{e704} 
and some predictions for RHIC energy can be attempted \cite{qs}. 

\vskip 12pt
{\bf 2.2 Intrinsic $\bfk_\perp$ in QCD factorization}
\vskip 8pt

A somewhat analogous approach has been discussed in 
Refs.~\cite{siv,noi1,dan,noi2,bl}; again, one starts from the leading twist,
collinear configuration scheme of Eq. (\ref{lt}), and generalizes it
with the inclusion of intrinsic transverse motion of partons in distribution
functions and hadrons in fragmentation processes:
\be
d\sigma = \sum_{a,b,c} f_{a/p}(x_1,\bfk_{\perp 1}) \otimes 
f_{b/p}(x_2, \bfk_{\perp 2}) \otimes
d\hat\sigma^{ab \to c \dots}(x_1, x_2, \bfk_{\perp 1}, \bfk_{\perp 2}) 
\otimes D_{h/c}(z, \bfk_{\perp h}) \>.
\label{ltgen}
\ee

The introduction of $\bfk_\perp$ and spin dependences opens up the way to
many possible spin effects; these can be summarized by new polarized
distribution functions,
\bea
\Delta^Nf_{q/\pup}  \!\!\! &\equiv& \!\!\!
\hat f_{q/\pup}(x, \bfk_{\perp})-\hat f_{q/\pdown}(x, \bfk_{\perp}) =  
\hat f_{q/\pup}(x, \bfk_{\perp})-\hat f_{q/\pup}(x, - \bfk_{\perp}) 
\label{delf1} \\
\Delta^Nf_{\qup/p}  \!\!\! &\equiv& \!\!\!
\hat f_{\qup/p}(x, \bfk_{\perp})-\hat f_{\qdown/p}(x, \bfk_{\perp}) =
\hat f_{\qup/p}(x, \bfk_{\perp})-\hat f_{\qup/p}(x, - \bfk_{\perp})  
\label{delf2}
\eea
and new polarized fragmentation function,
\bea 
\!\!\!\! \Delta^N D_{h/\qup} \!\!\!\! &\equiv& \!\!\!\!
\hat D_{h/\qup}(z, \bfk_{\perp}) - \hat D_{h/\qdown}(z, \bfk_{\perp}) =
\hat D_{h/\qup}(z, \bfk_{\perp})-\hat D_{h/\qup}(z, - \bfk_{\perp}) 
\label{deld1} \\
\!\!\!\! \Delta^N D_{\hup/q} \!\!\!\! &\equiv& \!\!\!\!
\hat D_{\hup/q}(z, \bfk_{\perp}) - \hat D_{\hdown/q}(z, \bfk_{\perp}) =
\hat D_{\hup/q}(z, \bfk_{\perp})-\hat D_{\hup/q}(z, - \bfk_{\perp}) 
\label{deld2} 
\eea
which have a clear meaning if one pays attention to the arrows denoting the
polarized particles. Details can be found, for example, in Ref. \cite{vill}.
All the above functions vanish when $k_\perp=0$ and are na\"{\i}vely $T$-odd.
The ones in Eqs. (\ref{delf2}) and (\ref{deld1}), when written in the helicity
basis, relate quarks of different helicities and are chiral-odd, while the 
other two are chiral-even.   

Similar functions have been introduced in the literature with different 
notations: in particular there is a direct correspondence \cite{bm} between 
the above functions and the ones denoted, respectively, by: $f_{1T}^\perp$
\cite{dp}, $h_1^\perp$ \cite{dan}, $H_1^\perp$ and 
$D_{1T}^\perp$ \cite{dp,jac}. For a comprehensive discussion and a better 
understanding the reader is urged to look at the lectures by P. Mulders in 
this school \cite{mul}.  
The function in Eq. (\ref{deld1}) is the Collins function \cite{col},
while that in Eq. (\ref{delf1}) was first introduced by Sivers \cite{siv}.

By inserting the new functions into Eq. (\ref{lt}), and keeping only 
leading terms in $k_\perp$ (see the comment at the end of this section), 
one obtains:
\bea
d\sigma^\uparrow - d\sigma^\downarrow \!\!\! &=& \!\!\! \sum_{a,b,c} \Bigl\{
\Delta^Nf_{a/\pup} (\bfk_{\perp}) \otimes f_{b/p}
\otimes
d\hat\sigma(\bfk_{\perp}) \otimes D_{\pi/c} \nonumber \\
\!\!\!&+& \!\!\!\! h_1^{a/p} \otimes f_{b/p} \otimes
\Delta\hat\sigma(\bfk_\perp) \otimes \Delta^N D_{\pi/c}(\bfk_\perp) 
\label{gen}\\  
\!\!\!&+& \!\!\!\! h_1^{a/p} \otimes \Delta^Nf_{\bup/p} (\bfk_{\perp}) \otimes
\Delta'\hat\sigma(\bfk_\perp) \otimes D_{\pi/c}(z) \Bigr\}\>, \nonumber
\eea
where the convolution now involves also a $\bfk_\perp$ integration
(we have explicitely shown the $\bfk_\perp$ dependences). The 
$\Delta\hat\sigma$'s denote polarized elementary interactions,
computable in pQCD: 
\bea
\Delta\hat\sigma &=& d\hat \sigma^{\aup b \to \cp d}  
- d\hat \sigma^{\aup b \to \cdown d} \label{dnn1} \\
\Delta'\hat\sigma &=& d\hat \sigma^{\aup \bup \to c d}  
- d\hat \sigma^{\aup \bdown \to c d} \> . \label{dnn2}
\eea
Notice that in the physical quantity given by Eq. (\ref{gen}) 
only products of even numbers of chiral-odd functions appear; it might be
interesting to consider also the similarity between Eq. (\ref{gen}) and
Eq. (\ref{htqs})

The above expressions have been used to successfully fit the E704 data,
either with the Sivers effect only \cite{noi1},
\be
d\sigma^\uparrow - d\sigma^\downarrow = \sum_{a,b,c}
\Delta^Nf_{a/\pup} (\bfk_{\perp}) \otimes f_{b/p}
\otimes d\hat\sigma(\bfk_{\perp}) \otimes D_{\pi/c} \>, \label{siv} 
\ee
or the Collins effect only \cite{noi2,bl}, 
\be
d\sigma^\uparrow - d\sigma^\downarrow = \sum_{a,b,c}
h_1^{a/p} \otimes f_{b/p} \otimes \Delta\hat\sigma(\bfk_\perp) \otimes 
\Delta^N D_{\pi/c}(\bfk_\perp) \>. \label{coll}
\ee

Some words of caution are necessary concerning the Sivers 
function $\Delta^Nf_{q/\pup}$, which is proportional to off-diagonal
(in helicity basis) expectation values of quark operators between
proton states \cite{col}: 
\be
\Delta^Nf_{a/\pup} \sim 
\langle p+|\, \overline\psi \gamma^+ \psi \, |p\, -\rangle \>.
\ee    

By exploiting the usual QCD parity and time-reversal properties for
free states one can prove the above quantity to be zero \cite{col}.
This might eliminate the Sivers function from the possible phenomenological
explanations of single spin asymmetries. A similar criticism might apply to 
the function defined in Eq. (\ref{delf2}). However, Sivers effect might be
rescued by initial state interactions, or by a new and subtle interpretation
of time reversal properties, discussed in the lecture by A. Drago \cite{dra}. 
 
Intrinsic $\bfk_\perp$ has been taken into account in Eqs. (\ref{gen}), 
(\ref{siv}) and (\ref{coll}) only ``minimally'': that is, in terms in which 
its neglect would lead to a zero value for $d\sigma^\uparrow 
- d\sigma^\downarrow$. In principle, it might be present in all terms: 
actually, according to Refs. \cite {wang,apa}, it should be taken into account 
in all distribution fuctions, where it leads to enhancing factors crucial 
for predicting the correct value of the unpolarized cross-section. We 
have not introduced these factors as they are spin independent and cancel 
in the ratio of cross-sections which defines the single spin asymmetry 
(\ref{spa}). 

\vskip 12pt\nd
{\bf 3. Fragmentation of a transversely polarized quark}   
\vskip 8pt

We have just seen that the Collins function, Eq. (\ref{deld1}), can explain 
the E704 data on $\pup \, p \to \pi \, X$ single spin asymmetry 
\cite{noi2,bl}; the natural question now is: are there other ways of 
accessing it, in order to get independent estimates of its size?

The answer to this question brings us to another of the single spin effects I 
listed above, namely the azimuthal asymmetry observed by HERMES and SMC 
collaborations in semi-inclusive DIS \cite{herm,smc}, 
$\ell \, \pup \to \ell \, \pi \, X$. Such asymmetries are directly related 
to the Collins function. In fact Eq. (\ref{deld1}) can be rewritten as:
\be
D_{h/\qup}(z, \bfk_\perp; \bfP_q) = \hat D_{h/q}(z, k_\perp) 
+ \frac 12 \> \Delta^ND_{h/\qup}(z, k_\perp) \> 
\frac{\bfP_q \cdot (\bfp_q \times \bfk_\perp)}
{|\bfp_q \times \bfk_\perp|}\>, \label{colfn} 
\ee
for a quark with momentum $\bfp_q$ and a transverse polarization 
vector $\bfP_q$ ($\bfp_q \cdot \bfP_q = 0$) which fragments into a hadron 
with momentum $\bfp_h = z\bfp_q + \bfk_\perp$ ($\bfp_q \cdot \bfk_\perp = 0$). 
$\hat D_{h/q}(z, k_\perp)$ is the unpolarized, $k_\perp$ dependent, 
fragmentation function. Parity invariance demands that the only component 
of the polarization vector which contributes to the spin dependent part
of $D$ is that perpendicular to the $q-h$ plane; in general one has:
\be
\bfP_q \cdot 
\frac{\bfp_q \times \bfk_\perp} {|\bfp_q \times \bfk_\perp|}
= P_q \sin\Phi_C \>, \label{colan}
\ee
where $P_q = |\bfP_q|$ and $\Phi_C$ is the Collins angle. When $P_q =1$ and 
$\bfP_q$ is perpendicular to the $q-h$ plane ($\bfP_q = \, \uparrow$, 
$-\bfP_q = \, \downarrow$) one has $P_q \sin\Phi_C = 1$. 

Eq. (\ref{colfn}) then implies the existence of a {\it quark analysing power}:
\be
A_q^h(z,k_\perp) \equiv \frac {\hat D_{h/\qup}(z,k_\perp) - 
\hat D_{h/\qdown}(z,k_\perp)} 
{\hat D_{h/\qup}(z,k_\perp) + \hat D_{h/\qdown}(z,k_\perp)} = 
\frac{\Delta^ND_{h/\qup}(z,k_\perp)}{2 \hat D_{h/q}(z,k_\perp)} \> \cdot
\ee

Eq. (\ref{ltgen}), keeping only leading terms in $k_\perp$, now gives
\be
d\sigma^\uparrow - d\sigma^\downarrow = \sum_q
f_{q/p} \otimes d\hat{\sigma} \otimes 
\Delta^N D_{\pi/c}(\bfk_\perp) \label{coldis}
\ee
and this results in a single spin asymmetry, which, in the $\gamma^*$-$p$
center of mass frame (where the pion intrinsic $\bfk_\perp$ coincides with
its observed $\bfp_T$, relatively to the colliding direction) reads 
\cite{noi4}:
\bea  
&& A^h_N(x,y,z,\Phi_C,p_T) =
 \frac{d\sigma^{\ell + p,\bfP \to \ell + h + X}
      -d\sigma^{\ell + p,-\bfP \to \ell + h + X}}
      {d\sigma^{\ell + p,\bfP \to \ell + h + X}
      +d\sigma^{\ell + p,-\bfP \to \ell + h + X}} \nonumber \\
&& = \frac{\sum_q e_q^2 \, h_1^{q/p}(x) \> \Delta^ND_{h/\qup}(z, p_T)}
{2\sum_q e_q^2 \, f_{q/p}(x) \> \hat D_{h/q}(z, p_T)} \>
\frac{2(1-y)} {1 + (1-y)^2} \> P \> \sin\Phi_C \>, \label{asym1} 
\eea
where $P$ is the transverse (with respect to the $\gamma^*$ direction)
proton polarization and I have introduced the usual DIS variables
\be
x = \frac{Q^2}{2p \cdot q}\>, \quad\quad y = \frac{Q^2}{sx} \>, 
\quad\quad z = \frac{p \cdot p_h}{p \cdot q} \>\cdot \label{var}
\ee
$p, q$ and $p_h$ are, respectively, the four-momenta of the proton, $\gamma^*$
and produced hadron. I have not indicated the $Q^2$ evolution dependence.  

We wonder how large the quark analysing power $A_q^h(z,p_T)$ can be. Such a 
question has been addressed in Ref. \cite{noi4}, where recent data on 
$A^\pi_N$ \cite{herm,smc} were considered. I refer to that paper
for all the details and only outline the main procedure here. 
Under some realistic assumptions and using isospin and charge 
conjugation invariance, Eq. (\ref{asym1}) gives ($i = +,-,0$):
\be
A^{\pi^i}_N(x,y,z,\Phi_C, p_T) = 
\frac{h_i(x)}{f_i(x)} \> A_q^\pi(z, p_T) \> 
\frac{2(1-y)} {1 + (1-y)^2} \> P \> \sin\Phi_C \label{aspi} 
\ee
where:
\bea
i &=& + : \quad h_+ = 4h_1^{u/p} \quad\quad 
f_+ = 4f_{u/p} +  f_{\bar d/p} \label{hf+} \\ 
i &=& - : \quad h_- =  h_1^{d/p} \quad\quad 
f_- =  f_{d/p} + 4f_{\bar u/p}  \label{hf-} \\ 
i = 0 : \quad h_0 &=& 4h_1^{u/p} + h_1^{d/p} \quad\quad 
f_0 = 4f_{u/p} + f_{d/p} + 4f_{\bar u/p} + f_{\bar d/p} \>. \label{hf0}
\eea
  
The $f$'s are the unpolarized distribution functions and the $h_1$'s
are the transversity distributions. Notice that the above equations
imply -- at large $x$ values -- $A^{\pi^+}_N \simeq A^{\pi^0}_N$ as 
observed by HERMES \cite{pi0}.

The measurable asymmetry (\ref{aspi}) depends on two unknown functions,
the tran\-sversity distribution and the quark analysing power, or Collins 
function. However, these functions depend on different variables (if we 
neglect the smooth $Q^2$ dependence induced by QCD evolution);
HERMES have an interesting experimental program according to
which, combining different measurements at different $z$, $x$ and $p_T$ 
values, separate evaluations of the two functions can be obtained \cite{kno}. 

Aiming at obtaining information on the size of $A_q^\pi$, one can use the only 
information one has on the transversity distribution, which is given by
the Soffer's inequality \cite{sof}
\be
|h_{1q}| \le \frac12 \, (f_{q/p} + \Delta q) \>, \label{sofb}
\ee
which, inserted into Eq. (\ref{aspi}), and comparing with SMC data 
\cite{smc}
\be
A_N^{\pi^+} \simeq -(0.10 \pm 0.06)\,\sin\Phi_C \>, \label{an+}
\ee
yields the significant lower bound for pion valence quarks:
\be
|A_q^\pi(\langle z \rangle, \langle p_T \rangle)| \>
\simorder \> (0.24 \pm 0.15)
\quad\quad \langle z \rangle \simeq 0.45 \>, \quad 
\langle p_T \rangle \simeq 0.65 \> \mbox{GeV}/c \>.
\label{res}
\ee  

A similar result is obtained by using HERMES data \cite{herm}, although their
proton transverse polarization is much smaller. Moreover, HERMES data are
taken at a much lower $Q^2$ value than SMC, and a correct comparison of
theoretical expressions with those data should take into account also higher 
twist contributions, which are not included in Eq. (\ref{asym1}). 
These lower bounds of the quark analysing power are anyway remarkably large. 

\vskip 12pt \nd
{\bf 4 Fragmentation of an unpolarized quark}
\vskip 8pt

Having dealt with the fragmentation of a transversely polarized quark
into a (spinless) pion, we are somewhat lead to consider the complementary
process, {\it i.e.} the possibility that an unpolarized quark fragments into 
a transversely polarized hadron \cite{multan,noi3}, which is, once again,
described by one of our 4 new functions brought to life by allowing for 
intrinsic transverse motion, Eq. (\ref{deld2}). This arises a sudden hope, 
that of being at last able of tackling in a consistent framework the thorny
problem of hyperon polarization \cite{lamdata,lamth}, which was on top of 
my initial list of hadronic single spin asymmetries.

In analogy to Eq. (\ref{colfn}) one can write
\be  
\hat D_{\hup/q}(z, \bfk_\perp; \bfP_h) = \frac 12 \> \hat D_{h/q}(z, k_\perp) 
+ \frac 12 \> \Delta^ND_{\hup/q}(z, k_\perp) \>  
\frac{\hat{\bfP}_h \cdot (\bfp_q \times \bfk_\perp)} 
{|\bfp_q \times \bfk_\perp|} \label{lamfn} 
\ee 
for an unpolarized quark with momentum $\bfp_q$ which fragments into  
a spin 1/2 hadron $h$ with momentum $\bfp_h = z \bfp_q + \bfk_\perp$ 
and polarization vector along the $\uparrow \> = \hat{\bfP}_h$ direction. 
$\Delta^ND_{\hup/q}(z, k_\perp)$ (denoted by $D_{1T}^\perp$ in 
Ref. \cite{multan}) is a new {\it polarizing fragmentation function}.

This indeed reflects into a possible polarization of hadrons inclusively
produced in the high energy interaction of unpolarized nucleons.
Let us recall that $\Lambda$ hyperons produced with $x_F \simorder 0.2$ 
and $p_T \simorder$ 1 GeV/$c$ in the collision of two unpolarized nucleons 
are polarized perpendicularly to the production plane, as allowed by parity 
invariance: 
\be 
P_\Lambda =  
\frac{d\sigma^{p N \to \Lup X} - d\sigma^{p N \to \Ldown X}} 
     {d\sigma^{p N \to \Lup X} + d\sigma^{p N \to \Ldown X}}  
\label{pol} 
\ee 
is found to be large and negative.

By taking into account intrinsic $\bfk_\perp$ in the hadronization  
process, and assuming that the factorization theorem holds also when  
$\bfk_\perp$'s are included, Eq. (\ref{ltgen}), one obtains
\bea 
&& d\sigma^{pN \to \Lambda X} \> P_\Lambda =  
 d\sigma^{pN \to \Lup X} - d\sigma^{pN \to \Ldown X} \label{phsch} \\  
&=& \!\!\!\! \sum_{a,b,c,d} f_{a/p}(x_1) \otimes  f_{b/N}(x_2)  
\otimes d\hat\sigma^{ab \to cd}(x_a, x_b, \bfk_\perp)  
\otimes \Delta^ND_{\Lup/c}(z, \bfk_\perp) \>. \nonumber  
\eea

In Ref. \cite{noi3} the above equation, together with a simple 
parameterization of the unknown polarizing fragmentation functions, was 
used to see whether or not one can fit the experimental data on $\Lambda$ 
and $\bar \Lambda$ polarization. Details can be found there. 
The data can be described with remarkable accuracy in all their features:
the large negative values of the $\Lambda$ polarization, the increase of
its magnitude with $x_F$, the puzzling flat $p_T \simorder 1$ GeV/$c$ 
dependence and the $\sqrt s$ apparent independence; data from $p$--$p$
processes are in agreement with data from $p$--$Be$ interactions and also the 
tiny or zero values of $\bar\Lambda$ polarization are well reproduced. 
The resulting functions $\Delta^ND_{\Lup/q}$ are very reasonable and 
realistic.  

However, at the moment of writing this contribution I have to add a 
necessary comment, due to new information \cite{comm}: in Ref. \cite{noi3} 
the data were supposed to be in the kinematical region of applicability of 
Eq. (\ref{phsch}), but a recent computation of the unpolarized cross-section 
in the same region shows that the contribution of the collinear pQCD 
factorization scheme, Eq. (\ref{lt}), gives only a few \% of the experimental 
value (at least for regions where both polarization and cross-section data 
are available, which is only a subset of the total region where data on
$P_\Lambda$ have been published and used). This might imply that Eq. 
(\ref{phsch}) cannot be compared with available data and that only higher 
energy and $p_T$ information could be used as a test of the scheme. 

It should also be noticed that a similar situation, regarding unpolarized 
cross-sections, holds for pion or photon productions, for which the pQCD 
calculations, even in the central rapidity region and at large $p_T$ values, 
can be a factor 100 smaller than data \cite{wang,ww,zfpbl}; in those cases 
the discrepancy is explained by the introduction of $\bfk_\perp$ effects in 
the distribution functions, Eq. (\ref{ltgen}): these give a large, spin 
independent, enhancing factor, which brings the cross-sections in agreement 
with data. Such factors would not alter the calculation of the $\Lambda$
polarization, as they cancel out in Eq. (\ref{pol}). However, it is too early 
to draw any definite conclusion, and a more detailed study is in progress.

\vskip 12pt \nd
{\bf 5 Comments and conclusions}
\vskip 8pt

I have considered single spin asymmetries in processes in which the 
short distance physics -- interactions between quarks and gluons -- should 
play a relevant role. As it often happens with spin, the experimental data
are very surprising; while the hard scattering spin dynamics is very simple,
almost trivial, experiments show rich and unexpected results, which have
to be understood.

We have discussed a general approach which couples pQCD dynamics, with 
new non perturbative information, along the lines of QCD factorization
theorem, which has been so successful in describing and predicting
many high energy processes. The final hope is that of having a consistent
phenomenological scheme in which the hard, perturbative partonic interactions
can be convoluted with new soft, non perturbative quantities; these
can be taken from some experiments and used in others. Their $Q^2$ evolution
has still to be studied, altough some progresses have recently been made
\cite{evol}.

The whole approach is summarized by Eqs. (\ref{ltgen}) and 
(\ref{delf1})-(\ref{deld2}) which show all its aspects, both the 
perturbative and non perturbative ones. We have applied it to the description
of single spin asymmetries which have been measured in 3 processes: 
$\pup \, p \to \pi \, X$, $\ell \, \pup \to \ell \, \pi \, X$ and 
$p \, p \to \Lup \, X$. Thus, Eq, (\ref{ltgen}) specifies to 
Eqs. (\ref{siv}), (\ref{coll}), (\ref{coldis}) or (\ref{phsch}).

These equations can easily describe the existing data, not only concerning
their numerical values, which are essentially reproduced by a proper 
parameterization of the new functions, but also concerning their general 
features, like $p_T$, $x_F$ and $\sqrt s$ dependences, which are more
related to the general scheme. The approach results strongly encouraged; 
however, at this stage, we are still far from being able to give actual 
predictions. The available experimental observation only allows at most to 
fix the general features of the new unknown, spin and $\bfk_\perp$ dependent 
functions, but not their precise form. Moreover, some of the data might 
not be safely in kinematical regions where pQCD dynamics dominates, and 
there might be corrections, difficult to control, from higher twist and 
higher order contributions. The problem of universality and QCD evolution 
of the new functions is still open and has just begun to be tackled. 

Nevertheless, I think that such a program is worth being pursued; more
experimental data are in the process of being obtained, from DESY (HERMES) 
and RHIC experiments, and more will be available in the near future at CERN
from COMPASS. Much more theoretical work is needed, to study the properties
of the new functions, to build models for them, and to combine theoretical
and experimental studies in order to give predictions.   
 
\bigskip{\bf Acknowledgements}

{\small I would like to thank the organizers for a beautifully organized 
school, with interesting subjects, continuous interest and pleasant 
atmosphere. The work I presented is the result of my collaboration with 
M. Boglione, D. Boer, U. D'Alesio and F. Murgia: without them it would
have been impossible.}

\end{document}